\documentclass[conference]{IEEEtran}
\IEEEoverridecommandlockouts

\usepackage{cite}
\usepackage{amsmath,amssymb,amsfonts}
\usepackage{algorithmic}
\usepackage{graphicx}
\usepackage{textcomp}
\usepackage{xcolor}
\usepackage{url}
\usepackage{booktabs}
\usepackage{adjustbox}
\usepackage{makecell}
\def\BibTeX{{\rm B\kern-.05em{\sc i\kern-.025em b}\kern-.08em
    T\kern-.1667em\lower.7ex\hbox{E}\kern-.125emX}}
\begin{document}

\title{Leveraging Large Language Models for Institutional Portfolio Management: Persona-Based Ensembles

\thanks{We would like to express our gratitude to The University of Tokyo Data Science School and The University of Tokyo Data Science Practicum (\url{https://dss.i.u-tokyo.ac.jp/}) for supporting this project.}
}
\author{\IEEEauthorblockN{Yoshia Abe}
\IEEEauthorblockA{\textit{Graduate School of Information Science and Technology} \\
\textit{The University of Tokyo}\\
Tokyo, Japan \\
y-abe@isi.imi.i.u-tokyo.ac.jp (0009-0007-0885-8852)}
\and
\IEEEauthorblockN{Shuhei Matsuo}
\IEEEauthorblockA{\textit{Graduate School of Information Science and Technology} \\
\textit{The University of Tokyo}\\
Tokyo, Japan \\
matsuo@isi.imi.i.u-tokyo.ac.jp
}
\and
\IEEEauthorblockN{Ryoma Kondo}
\IEEEauthorblockA{\textit{Graduate School of Information Science and Technology} \\
\textit{The University of Tokyo}\\
\textit{The Canon Institute for Global Studies}\\
Tokyo, Japan \\
kondor@g.ecc.u-tokyo.ac.jp
}
\and
\IEEEauthorblockN{Ryohei Hisano}
\IEEEauthorblockA{\textit{Graduate School of Information Science and Technology} \\
\textit{The University of Tokyo}\\
\textit{The Canon Institute for Global Studies}\\
Tokyo, Japan \\
hisanor@g.ecc.u-tokyo.ac.jp}
}
\maketitle
\begin{abstract}

Large language models (LLMs) have demonstrated promising performance in various financial applications, though their potential in complex investment strategies remains underexplored. To address this gap, we investigate how LLMs can predict price movements in stock and bond portfolios using economic indicators, enabling portfolio adjustments akin to those employed by institutional investors. Additionally, we explore the impact of incorporating different personas within LLMs, using an ensemble approach to leverage their diverse predictions. Our findings show that LLM-based strategies, especially when combined with the mode ensemble, outperform the buy-and-hold strategy in terms of Sharpe ratio during periods of rising consumer price index (CPI). However, traditional strategies are more effective during declining CPI trends or sharp market downturns. These results suggest that while LLMs can enhance portfolio management, they may require complementary strategies to optimize performance across varying market conditions.
\end{abstract}
\section{Introduction}
Large language models (LLMs) exhibit a wide range of capabilities that extend beyond traditional natural language processing tasks. In the financial sector, LLMs are increasingly employed to enhance decision-making and improve operational efficiency. For example, BlackRock has explored innovative methods for classifying companies using LLMs \cite{Vamvourellis2023}. Similarly, \cite{Trajanoska2023} used LLMs to extract structured environmental, social, and governance (ESG) data from sustainability reports to build a knowledge graph that facilitates deeper analysis of corporate sustainability practices. LLMs have also been employed to detect accounting fraud in the Management Discussion and Analysis sections of 10-K reports, surpassing existing benchmark models \cite{Bhattacharya2024}. These examples and others \cite{Nie2024} illustrate the expanding role of LLMs in finance, though further exploration is required to understand their full potential in more complex investment strategies. 

Narrowing the focus to investment-related applications, LLMs have shown promising results in various aspects of portfolio management. For instance, \cite{Ko2024} demonstrated that assets selected by GPT models outperform randomly chosen assets in terms of diversification and average return. Although GPT excels in stock selection, optimization models perform better at portfolio allocation, prompting researchers to propose a strategy that combines the strengths of both \cite{Romanko2023}. Additionally, \cite{Cheng2024} showed that GPT models can create economically explainable factors based solely on their knowledge base, leading to the development of a new model based on these factors. However, despite these advances, most research remains focused on portfolio management at the individual stock level, with limited attention paid to institutional investors, who often manage portfolios at a more granular and complex level.

In finance, understanding investor attitudes is crucial, as decisions are shaped by beliefs, values, and preferences. Individual investors, for example, often make short-term decisions \cite{Barber2000}, whereas institutional investors typically adopt a long-term perspective and are less influenced by behavioral biases \cite{Skiba2017}. Additionally, research shows that gender differences influence risk perception and management among investment professionals, with women placing greater emphasis on risk reduction, particularly in extreme scenarios \cite{Olsen2001}. Modeling these diverse investment attitudes through LLMs could provide a powerful tool for personalizing financial strategies, especially for institutional investors who must account for various needs and behaviors when managing large portfolios.

Interestingly, just as individual investors vary in their preferences and performance, LLMs exhibit significant variation in their outputs depending on the specified persona \cite{Xu2023,Tseng2024}. For example, \cite{Wu2023} introduced DR-CoT prompting, in which LLMs use personas to mimic the diagnostic reasoning processes of medical professionals. Similarly, \cite{Fu2023} employed personas (e.g., buyer, seller, and critic) in a gaming environment to evaluate whether LLMs can autonomously enhance their strategies through iterative interactions and mutual feedback. While these personas have proven effective in other fields, their application in finance, particularly in replicating investor attitudes or investment strategies, remains underexplored. Leveraging this capability could provide novel insights into portfolio management, especially in institutional settings, where nuanced or mixed decision-making is critical.

Building on this foundation, we explore the task of inputting economic indicator data into LLMs to predict the price movements of a portfolio consisting of stocks and government bonds, adjusting positions based on the obtained predictions in a manner similar to the portfolio management of institutional investors. Additionally, we investigate how the performance of LLMs varies depending on the specified persona, applying these differences in an ensemble approach to construct the final portfolio. Because the effectiveness of investment strategies can vary depending on the testing period, we compare LLM-based portfolio management strategies with baseline models in detail, analyzing the conditions under which LLMs are most effective. Furthermore, we qualitatively examine the reasoning behind the LLM's predictions, enabling a deeper understanding of the decision-making process and the key information it focuses on.

We find that LLM-based predictions, particularly when combined with the ensemble approach, detect market declines well. Moreover, LLM-based investment strategies outperform the buy-and-hold strategy in terms of the Sharpe ratio during periods of rising consumer price index (CPI), whereas buy-and-hold performs better during a declining CPI. For other metrics, such as return, volatility, and maximum drawdown, different strategies tend to perform best depending on market conditions. Additionally, LLM-based strategies generally respond effectively to sharp market declines by reducing positions, though traditional strategies can offer better protection during rapid downturns.

The contributions of this study are as follows: 
\begin{enumerate} 
\item Prompts that enable LLMs to manage portfolios in line with institutional investor settings are designed. 
\item Differences in the performance of portfolio strategies based on LLM personas are investigated and leveraged via ensemble methods. 
\item Periods when LLM-based strategies excel are quantitatively analyzed and the LLM-generated rationales for each persona are qualitatively analyzed. 
\item LLM-based strategies are shown outperform traditional methods in Sharpe ratio during rising CPI trends. 
\end{enumerate}

\section{Related Work}

As mentioned above, LLMs have seen widespread application in finance in recent years. The related research can be broadly categorized into three areas: financial concept comprehension, academic applications, and investment decision-making \cite{Li2023}. Research on investment decision-making can be further divided into studies focused on individual investors and those focused on institutional investors, with most current studies concentrating on the former. One of the few studies focusing on institutional investors is \cite{Cheng2024}, which used GPT-4 to generate high-return equity investment factors, achieving an annualized return of up to 88\% and a Sharpe ratio of 2.46, significantly outperforming traditional models. However, their approach is limited to stock prices and does not predict price movements in a portfolio consisting of both stocks and government bonds, nor does it adjust positions based on these predictions in a manner consistent with institutional portfolio management.

We also review prompt engineering, which is widely recognized as a crucial step in enhancing the capabilities of LLMs. The studies \cite{Dhar2023,Xu2023} demonstrated that specifying a persona; providing concrete examples of investment strategies; and clearly defining the objective, output content, and format significantly improve response quality. In particular, studies on personas have shown that adjusting the explanation level based on the persona of the intended audience can be effective \cite{Yue2023}. However, the focus was on financial concept comprehension, not comparing investment performance when the persona of the institutional investor was altered.

\begin{figure*}[ht]
\centering
\includegraphics[width=1.6\columnwidth]{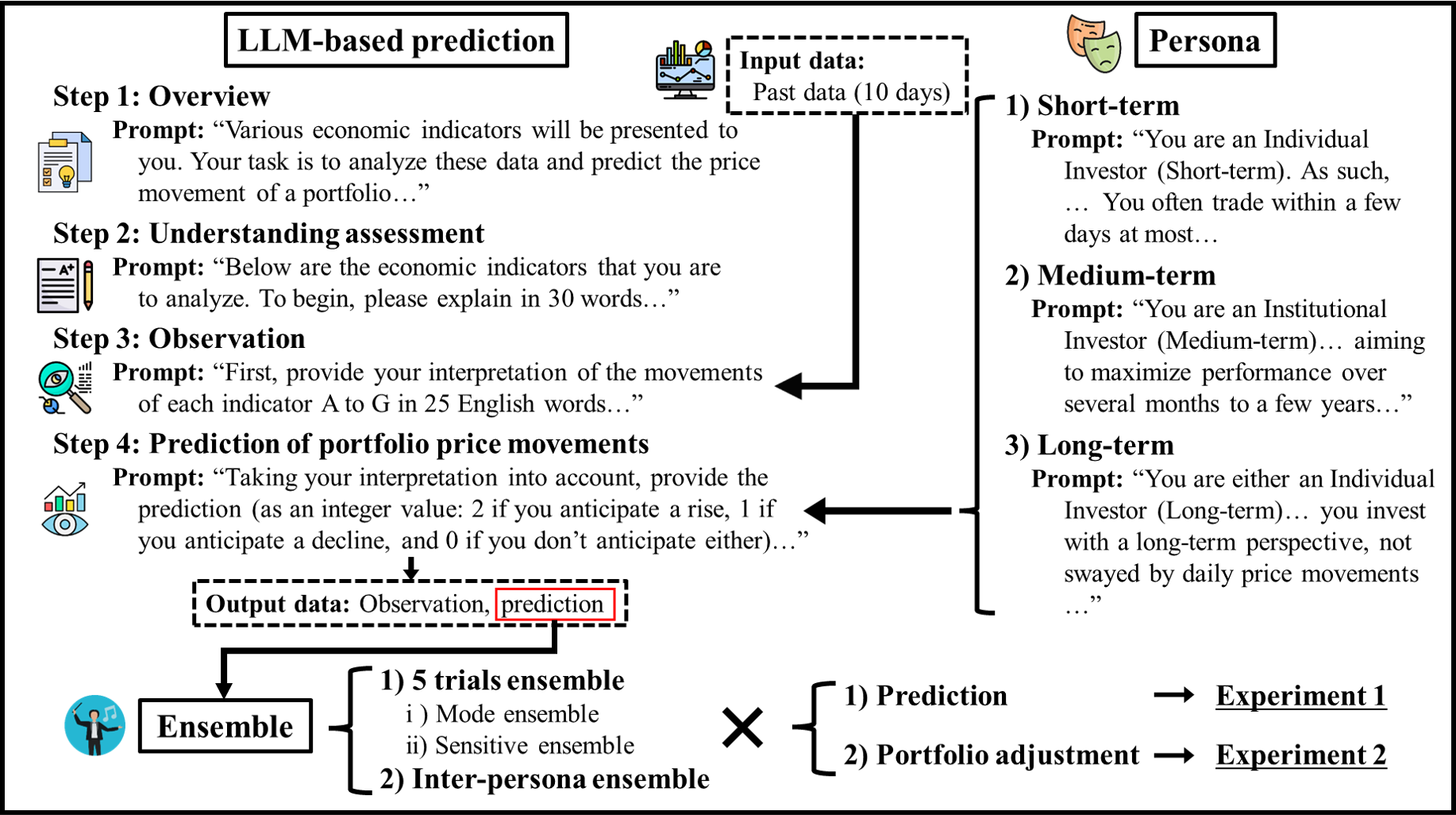}
\caption{Overview of our approach.}
\label{fig:concept}
\end{figure*}
\section{Methods}
\label{sec:methods}

\subsection{Task Definition}
\label{subsec:task_definition}
We investigate the ability of LLMs to first predict price movements in a portfolio consisting of stocks and government bonds, and then adjust positions based on these predictions in a manner similar to institutional portfolio management. The portfolio is composed of 40\% US equities and 60\% US bonds. The LLM is tasked with predicting whether the portfolio value will rise or fall by more than 2\% within the next 5 days based on data from the previous 10 days for the following seven numerical data indicators:

\begin{enumerate}
    \setlength{\parskip}{0cm} 
    \setlength{\itemsep}{0cm} 
    \item [A:] 40\% Stock, 60\% Bond Portfolio (Return)
    \item [B:] US Stocks (Futures, Return)
    \item [C:] US 5-year Interest Rate
    \item [D:] US 30-year Interest Rate
    \item [E:] US Interest Rate Spread, 30--10 year
    \item [F:] Volatility Index (VIX)
    \item [G:] US Dollar Index
\end{enumerate}

The output prediction is a three-class classification task, where ``0,'' ``1,'' and ``2'' indicate holding, falling, and rising, respectively. While various LLMs are available, we use GPT-4 (gpt-4-0613) \cite{OpenAI2023}, which is known for its strong performance across many tasks.

\subsection{Prompt Design}

The performance of LLMs varies significantly depending on prompt design, which has led to the development of various prompting methods \cite{OpenAI2024a, OpenAI2024b}. Table \ref{tab:prompting} lists common prompting methods. We use methods 1--9 to minimize the number of interactions with LLMs. The remaining methods require substantially more tokens and interactions with the LLMs, and hence we leave these for future work.

The following sequence of prompts was used in the experiment: First, an overview of the task and relevant considerations is provided as a system attribute prompt. Next, the LLM is queried to assess its understanding of various economic indicators. Then, numerical data from the past 10 days for various economic indicators are input, and the model is asked to interpret the trends. Finally, based on these observations and interpretations, the model is tasked with predicting the rise or fall of a portfolio. Our approach is illustrated in Fig. \ref{fig:concept}. The complete version of the prompt is available on our GitHub page\footnote{\url{https://github.com/YoshiaAbe/llm_based_portfolio_management}}.
\begin{table*}[htbp]
    \centering
    \caption[Prompting]{Prompting methods. Check mark \checkmark indicates that the method was used in the experiments in this study.}
    \begin{tabular}{c|l|c}
        No. & Name and Explanation & Used \\
        \hline
        1 & Use clear and concrete instructions \cite{OpenAI2024a,OpenAI2024b}.& \checkmark \\
        2 & Use delimiters such as \#\#\# \cite{OpenAI2024a,OpenAI2024b}.& \checkmark \\
        3 & Specify the output length, style, and similar factors in detail. \cite{OpenAI2024a,OpenAI2024b}& \checkmark \\
        4 & Specify the output format, e.g., in JSON \cite{OpenAI2024a,OpenAI2024b}. & \checkmark \\
        5 & Use "Refrain from ..." instead of "Don't do ..." \cite{OpenAI2024a}.& \checkmark \\
        6 & Specify the persona that you want the LLM to behave as \cite{OpenAI2024b,Xu2023}. & \checkmark \\ 
        7 & Ask the model to output its thoughts before its conclusion \cite{OpenAI2024b}. & \checkmark \\ 
        8 & Chain-of-Thought \cite{Wei2023}, Least-to-Most \cite{Zhou2023}: decompose the task into sub-tasks. & \checkmark \\ 
        9 & Zero-shot Chain-of-Thought \cite{Kojima2023}: tell the LLM to ``think step-by-step."  & \checkmark \\
        10 & Make the LLM evaluate whether it met the specified instruction \cite{OpenAI2024b}. &  \\ 
        11 & EchoPrompt \cite{Mekala2024}: make the LLM rephrase the question before answering. & \\
        12 & Few-shot Prompting \cite{Brown2020}: show the LLM examples of correct answers. &  \\
        13 & Contrastive Chain-of-Thought \cite{Chia2023}: show the LLM examples of incorrect reasoning. & \\
        14 & Self-consistency \cite{Wang2023}: the final answer is the majority choice among multiple outputs.  &  \\
        15 & Self-refine \cite{Madaan2023}: make the LLM provide feedback about itself and iteratively refine its output. &  \\
        16 & Tree of Thoughts \cite{Yao2023}: Manage and explore a chain of thought in a tree structure. & \\
        \hline
    \end{tabular}
    \label{tab:prompting}
\end{table*}

\section{Experiments}

\subsection{Overview of the Experimental Procedure}
We use GPT-4 (gpt-4-0613), which was trained up to September 2021, in all of our experiments. To avoid data leakage and properly evaluate the predictive performance of the LLM, we focused on the period from October 2021 to January 2024, covering 593 weekdays. Moreover, we repeated the predictions five times to account for variability in the LLM outputs.

We conducted two experiments. Because LLM inference capabilities are heavily influenced by the persona specified in the prompt \cite{Xu2023} and investment strategies often vary based on individual beliefs, values, and aims, we sought to leverage this heterogeneity. In the first experiment, we examined how different personas affect prediction accuracy. The performance was evaluated across multiple trials using metrics such as accuracy, precision, recall, and F1-score. Additionally, we assessed the performance of several ensemble methods to determine whether the predictions from different personas could be effectively integrated.

In the second experiment, we developed an investment strategy based on the LLM's price movement predictions and compared it with a baseline strategy that did not use an LLM. The evaluation period was divided into months, each characterized by economic indicator data (US CPI Total), and we determined the periods over which the LLM-based strategies performed better.

\subsection{Experiment 1: Impact of Personas and Ensembles on Prediction Accuracy}

We investigated how different personas affect prediction accuracy and evaluated performance based on the time span of their investments using the following three persona conditions:
\begin{enumerate} 
\setlength{\parskip}{0cm} 
\setlength{\itemsep}{0cm} 
\item \textbf{Short term}: An individual investor who trades over a span of several days, with limited knowledge of investment and challenges in risk management. 
\item \textbf{Medium term}: An institutional investor trading over several months to a few years, with extensive knowledge of investment and robust risk management capabilities. 
\item \textbf{Long term}: Both individual and institutional investors who operate with a long-term perspective, spanning 20 to 30 years. 
\end{enumerate}

Additionally, we considered the following two ensemble methods to account for the variability in LLM predictions:
\begin{enumerate} 
\setlength{\parskip}{0cm} 
\setlength{\itemsep}{0cm} 
\item \textbf{Mode}: The final prediction is the class with the most votes across five trials. In the case of a tie, the class with the smaller number is prioritized. 
\item \textbf{Sensitive}: If Class 2 (rise) or Class 1 (fall) appears in any of the five trials, that class is chosen as the final prediction. If both Class 1 and Class 2 are present, the class with the most votes is selected. In the event of a tie, Class 0 is chosen. We call this method ``sensitive” because it is more likely to predict a rising or falling market than the mode method.
\end{enumerate}

\subsection{Experiment 2: Adaptive Investment Strategies with LLM Predictions}

In the second experiment, we evaluated the performance of investment strategies based on the predictive outputs from Experiment 1. Our strategy involved adjusting the position size of stocks and government bonds, ranging from $0.0$ to $1.0$, in increments of $0.2$. The investment period spanned 593 weekdays, from October 2021 to January 2024, as in Experiment 1, with the position starting at $1.0$ on the first day. For evaluation purposes, and to ensure sufficient data within each month, we focused on the 26 months between November 2021 and December 2023, excluding the initial and final months of the evaluation period used in Experiment 1.

We used the following actions for our investment strategy:
\begin{enumerate}
\setlength{\parskip}{0cm} 
\setlength{\itemsep}{0cm} 
\item \textbf{Pattern 1}: Binary actions (increase or decrease position size); 
\item \textbf{Pattern 2}: Ternary actions (increase, decrease, or maintain position size); 
\item \textbf{Pattern 3}: Ternary actions (increase, decrease, or maintain position size), with adjustments.
\end{enumerate}

In the first pattern, the position size decreases when the predicted class is 1 and increases when the predicted class is 0 or 2. Pattern 2 decreases the position size when the predicted class is 1, increases it when the predicted class is 2, and remains unchanged when the predicted class is 0. The final pattern, pattern 3, is similar to pattern 2 but differs in that if there is no change in position size for the past $d_{\mathrm{flat}}$ days, the position is gradually increased to $1.0$ in steps of $0.2$. In this study, $d_{\mathrm{flat}}$ was fixed at 5 days.

For comparison, we evaluated our strategy against the following baselines:
\begin{enumerate} 
\setlength{\parskip}{0cm} 
\setlength{\itemsep}{0cm} 
\item \textbf{Buy-and-hold}: This strategy maintains a constant position of 1.0 starting from the first day and continuing throughout the period.
\item \textbf{Continuous movement (CM)}: This strategy tracks indicator A (40\% stock, 60\% bond portfolio return values) over the past $d_{\mathrm{window}}$ days. If the value rises continuously for $d_{\mathrm{continuity}}$ days, the position is increased; if it falls continuously, the position is decreased. In this experiment, $d_{\mathrm{window}}$ was fixed at 10 days and two values of $d_{\mathrm{continuity}}$ were tested: 2 and 3 days (CM(D2) and CM(D3), respectively).
\item \textbf{Regression (RG)}: This strategy also observes indicator A over the past $d_{\mathrm{window}}$ days and performs a linear regression with the days as the explanatory variable and indicator A values as the dependent variable. If the slope of the regression line is positive, the position is increased; if it is negative, the position is decreased. However, if the absolute value of the slope is below $s_{\mathrm{threshold}}$, the position is not adjusted. In this experiment, $d_{\mathrm{window}}$ was fixed at 10 days and two values of $s_{\mathrm{threshold}}$ were tested: 0.001 and 0.0005 (RG(S10) and RG(S5), respectively). 
\end{enumerate}

We use the following four metrics to evaluate the portfolio management strategies: 

\textbf{Return:}
The cumulative return $R_{\mathrm{cumul}}$ over period $T$ is given by \( R_{\mathrm{cumul}} = \{\sum_{i}^{n}(1 + p_i r_i)\}- 1\), where $p_i$ and $r_i$ represent the position and return on the i-th day, respectively. The return metric is adjusted by dividing it by the average position size: \( \frac{1}{\bar{p}}R_{\mathrm{cumul}} \). 

\textbf{Volatility:}
The volatility $V$ over period $T$ is the standard deviation of $p_i r_i$, scaled by the square root of the number of trading days: \( V = \sqrt{\frac{1}{n}{\sum_{i}^{n}\{(p_i r_i) - \overline{p_i r_i}\}^2}} \sqrt{n}\). The volatility metric is adjusted by dividing it by the average position size: \( \frac{1}{\bar{p}}V\). 

\textbf{Maximum drawdown:}
The maximum drawdown $D_{\mathrm{max}}$ during period $T$ is the largest decline in asset value from a previous peak. Let $R_{\mathrm{cumul}, i}$ represent the cumulative return up to day $i$. The drawdown $D_i$ is calculated as \( D_i = R_{\mathrm{cumul}} - \max (R_{\mathrm{cumul}, 1}, R_{\mathrm{cumul}, 2}, \cdots ,R_{\mathrm{cumul}, i})\). The maximum drawdown is the lowest value of $D_i$ over $n$ days, multiplied by $-1$: \( D_{\mathrm{max}} = - \min (D_1, D_2, \cdots, D_n) \). The maximum drawdown metric is adjusted by dividing it by the average position size: \( \frac{1}{\bar{p}}D_{\mathrm{max}}\). 

\textbf{Sharpe ratio:}
The Sharpe ratio $S$ for period $T$ is calculated as \( \frac{1}{V}R_{\mathrm{cumul}}\) and is used as the evaluation metric.

\section{Results}
\subsection{Results of Experiment 1: Impact of Personas and Ensembles on Prediction Accuracy}
\begin{figure}[htbp]
\centering
\includegraphics[width=\columnwidth,trim=0 0 0 0,clip]{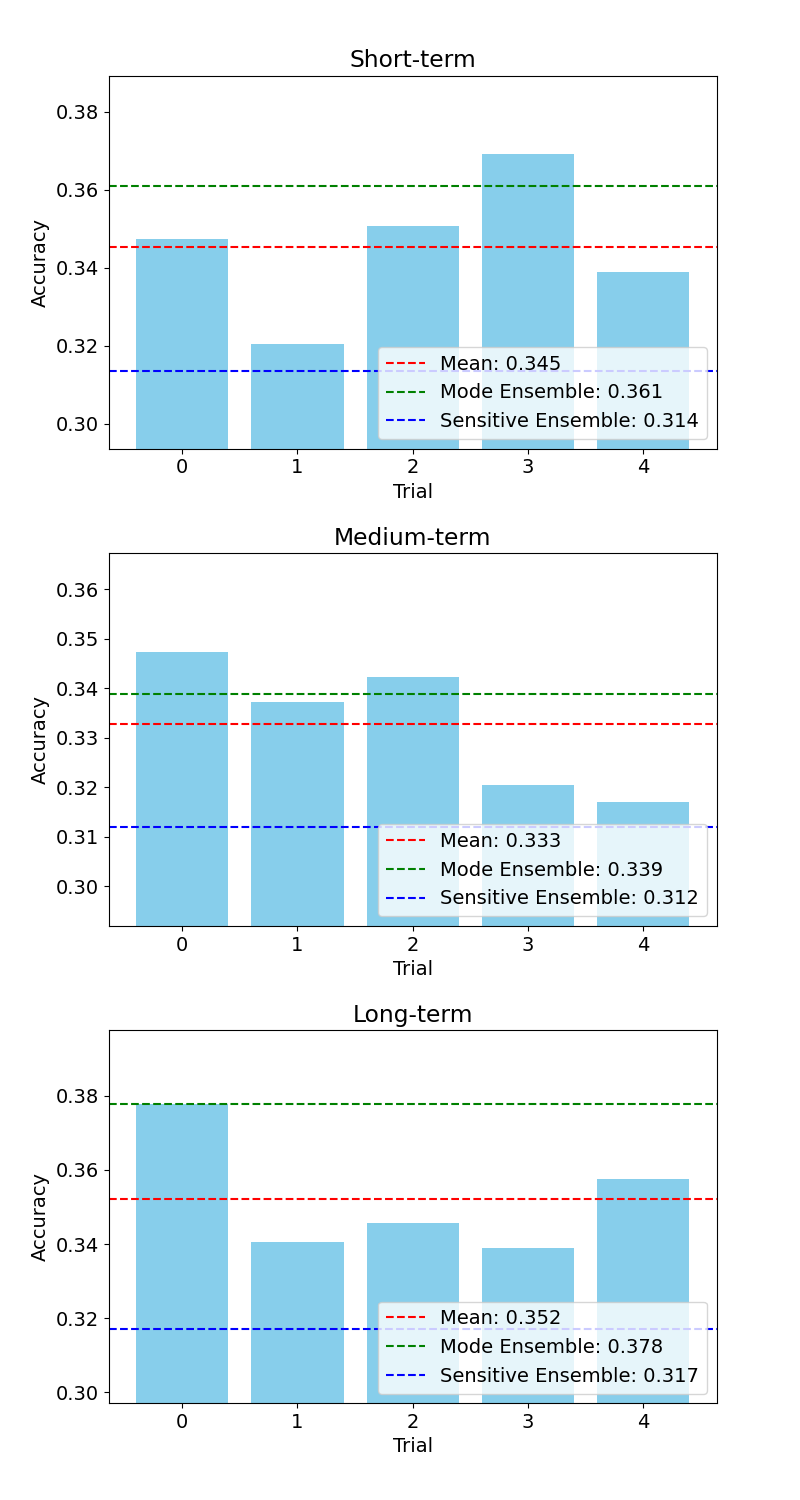}
\caption{Accuracy of five trials for each persona condition (short, medium, or long term)}
\label{fig:exp1_accuracy_individual_personas}
\begin{minipage}[c]{0.85\linewidth}
\end{minipage}
\end{figure}

First, we compare the predictive accuracy of each persona in Fig.~\ref{fig:exp1_accuracy_individual_personas}. The average accuracies across five trials for the short, medium, and long personas were 0.345, 0.333, and 0.352, respectively. When applying the mode ensemble within the same persona across trials, the accuracy improved to 0.361, 0.339, and 0.378, respectively. In contrast, the sensitive ensemble resulted in lower accuracy values of 0.314, 0.312, and 0.317, respectively. For all three personas, the mode ensemble consistently improved performance compared with the average values, whereas the sensitive ensemble caused performance to decline.

Second, we examine the performance of ensembles across different persona. For each persona, we generated prediction results using the mode or sensitive ensemble across the five trials. We then ensembled these predictions across the three personas, using either the mode or sensitive method, resulting in a total of four patterns (2×2). The results are presented in Table \ref{tab:exp1_accuracy_inter-persona_ensemble}. Among the four patterns, the highest accuracy of 0.366 was achieved by applying the mode ensemble across both the five trials and the three personas. The highest individual accuracy was obtained using the mode ensemble for the five trials of the long-term persona, which resulted in an accuracy of 0.378, as shown in the bottom panel of Fig.~\ref{fig:exp1_accuracy_individual_personas}.

To further evaluate the predictive accuracy of the LLMs, we report the precision, recall, F1-score, correct counts, and predicted counts for each class in Table \ref{tab:exp1_long-term_mode-ensemble_df-report}. The results indicate that, while approximately half of the correct labels fall into class 0, the LLM tends to predict class 1 more frequently. Given that the chance level of accuracy when randomly selecting a class from 0, 1, or 2 with a uniform distribution is 0.333, it is clear that the LLM’s predictions outperform chance. Notably, the mode ensemble contributes to an improvement in overall accuracy.

\begin{table}[htbp]
    \centering
    \caption{Accuracy of inter-persona ensembles. Rows and columns indicate the ensemble method applied across the three personas and within each persona, respectively.}
    \begin{tabular}{c||c|c|c|c|c}
        Inter-Persona Ensemble Method & Mode & Sensitive \\
        \hline
        Mode (across three personas) & 0.366 & 0.324 \\
        Sensitive (across three personas) & 0.319 & 0.307 \\
        \hline
    \end{tabular}
    \label{tab:exp1_accuracy_inter-persona_ensemble}
\end{table}
\begin{table}[htbp]
    \centering
    \caption{Precision, recall, F1-score, correct counts, and predicted counts for each class when using the mode ensemble for the five trials of the long-term persona.}
    \begin{tabular}{c||c|c|c|c|c}
        Class & Precision & Recall & F1-score & Corr. Cnt. & Pred. Cnt. \\
        \hline
        0 & 0.551 & 0.327 & 0.411 & 281.0 & 167.0 \\
        1 & 0.380 & 0.570 & 0.456 & 172.0 & 258.0 \\
        2 & 0.202 & 0.243 & 0.221 & 140.0 & 168.0 \\
        
        \hline
    \end{tabular}
    \label{tab:exp1_long-term_mode-ensemble_df-report}
\end{table}

\begin{figure}[htbp]
\centering
\includegraphics[width=\columnwidth,trim=0 0 0 0,clip]{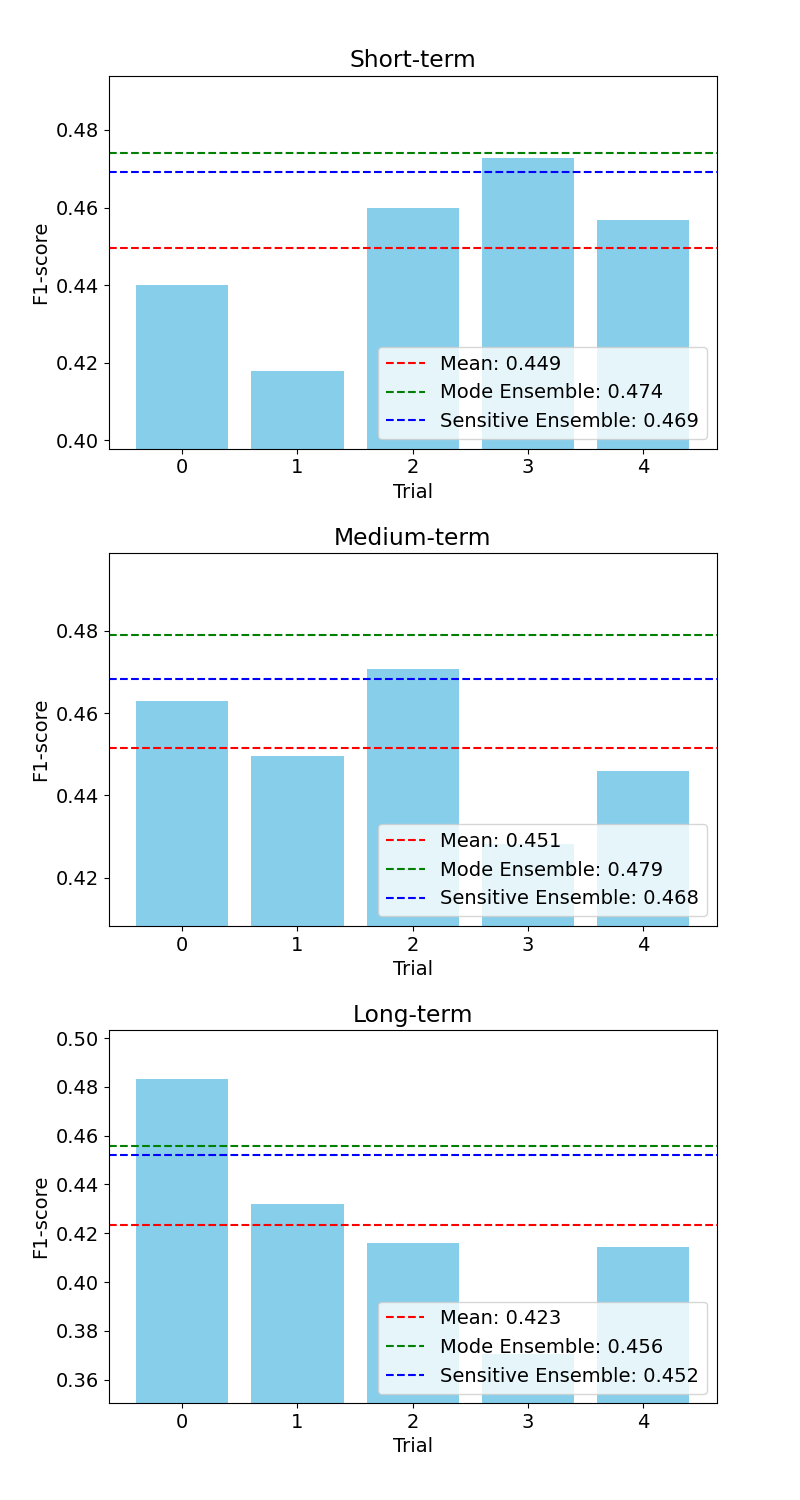}
\caption{F1-score (Class 1) of five trials for each persona condition (short term, medium term, and long term)}
\label{fig:exp1_f1score_individual_personas}
\begin{minipage}[c]{0.85\linewidth}
\end{minipage}
\end{figure}

For institutional investors, a model that predicts declines (class 1) with high precision and recall is particularly desirable to avoid significant losses. Therefore, we further investigated the F1-score for class 1, as shown in Fig. \ref{fig:exp1_f1score_individual_personas}. The average F1-scores across five trials for the short-, medium-, and long-term personas were 0.449, 0.451, and 0.423, respectively. When applying the mode ensemble within the same persona across trials, the scores improved to 0.474, 0.479, and 0.456, respectively. The sensitive ensemble yielded scores of 0.469, 0.468, and 0.452, respectively. Unlike accuracy,  the F1-scores of both the mode and sensitive ensembles were better than the average values.

The F1-score performance when using ensembles across different personas is presented in Table \ref{tab:exp1_f1score_inter-persona_ensemble}. The highest F1-score (0.484) was achieved by applying the mode ensemble across the five trials, followed by a mode ensemble across the three personas. The precision, recall, F1-score, correct counts, and predicted counts for each class in this case are listed in Table \ref{tab:exp1_mode_mode-ensemble_df-report}. Notably, while maintaining a precision of 0.378, the recall reached 0.674.

On the basis of these results, we conclude that the mode ensemble method, which uses majority voting for the final predictions, improves both accuracy and F1-score. Moreover, applying the mode ensemble across personas proves to be the most effective approach when focusing on the F1-score for predicting a declining market, which is critical for institutional investors. 

\begin{table}[htbp]
    \centering
    \caption{F1-score (Class 1) comparison of inter-persona ensemble. Rows and columns indicate the ensemble method applied across the three personas and within each persona, respectively.}
    \begin{tabular}{c||c|c|c|c|c}
        Inter-Persona Ensemble Method &  Mode & Sensitive \\
        \hline
        Mode (across three personas) & 0.484 & 0.466 \\
        Sensitive (across three personas) & 0.477 & 0.461 \\
        \hline
    \end{tabular}
    \label{tab:exp1_f1score_inter-persona_ensemble}
\end{table}
\begin{table}[htbp]
    \centering
    \caption{Precision, recall, F1-score, correct counts, and predicted counts for each class when using the mode ensemble for the five trials, followed by the mode ensemble across the three personas.}
    \begin{tabular}{c||c|c|c|c|c}
        Class & Precision & Recall & F1-score & Corr. Cnt. & Pred. Cnt. \\
        \hline
        0 & 0.556 & 0.263 & 0.357 & 281.0 & 133.0 \\
        1 & 0.378 & 0.674 & 0.484 & 172.0 & 307.0 \\
        2 & 0.176 & 0.193 & 0.184 & 140.0 & 153.0 \\

        \hline
    \end{tabular}
    \label{tab:exp1_mode_mode-ensemble_df-report}
\end{table}

\subsection{Results of Experiment 2: Adaptive Investment Strategies with LLM Predictions}
We first examine the Sharpe ratio to evaluate overall performance, balancing profit and risk. Table \ref{tab:strategies_comparison_with_mode-mode_in_sharp-ratio} presents the monthly Sharpe ratios for each strategy during the 26-month evaluation period. For each month, the best- and worst-performing strategies are highlighted in bold and underlined, respectively. Strategies that outperform the buy-and-hold strategy are indicated in red. The ``nan" entries indicate periods where the Sharpe ratio could not be calculated due to zero volatility. 

The LLM-based strategies (patterns 1, 2, and 3) used the prediction gained after the mode ensemble across the five trials, followed by the mode ensemble across the three personas. The red text reveals that these LLM-based strategies outperformed the buy-and-hold strategy in some periods but underperformed in others. Furthermore, examining the bold and underlined entries reveals that, for certain months, a strategy could be either the best or the worst performer among all strategies.

\begin{table*}[h]
\centering
\caption{Sharpe ratios for each strategy during the evaluation period. For each month, the best-performing strategy is highlighted in bold, whereas the worst one is underlined. Strategies that outperformed the buy-and-hold strategy are indicated in red.}
\small 
\begin{tabular}{lrrrrrrrr}
Year-Month & Buy-and-hold & Pattern 1 & Pattern 2 & Pattern 3 & CM(D2) & CM(D3) & RG(S10) & RG(S5) \\
2021-11 & \underline{0.097} & \textcolor{red}{0.280} & \textcolor{red}{\textbf{0.397}} & \textcolor{red}{\textbf{0.397}} & \textcolor{red}{0.108} & \underline{0.097} & \underline{0.097} & \textcolor{red}{0.172} \\
2021-12 & \textbf{1.003} & 0.559 & 0.424 & \underline{0.353} & 0.574 & 0.736 & \textbf{1.003} & 0.709 \\
2022-01 & -1.421 & -1.583 & -1.547 & -1.583 & \underline{-2.013} & -1.731 & -1.421 & \textcolor{red}{\textbf{-0.804}} \\
2022-02 & \textbf{-0.540} & -1.092 & \underline{-1.489} & \underline{-1.489} & -1.446 & -0.690 & -0.565 & -0.860 \\
2022-03 & -0.196 & \textcolor{red}{\textbf{-0.051}} & -0.306 & -0.306 & -0.731 & -0.556 & -0.614 & \underline{-0.861} \\
2022-04 & -1.826 & \textcolor{red}{-1.432} & \textcolor{red}{-1.432} & \textcolor{red}{-1.432} & \textcolor{red}{-1.391} & \textcolor{red}{\textbf{-1.167}} & \underline{-1.943} & -1.847 \\
2022-05 & \underline{0.136} & \textcolor{red}{0.922} & \textcolor{red}{0.404} & \textcolor{red}{0.404} & \textcolor{red}{0.266} & \textcolor{red}{\textbf{1.188}} & \textcolor{red}{0.147} & \textcolor{red}{0.544} \\
2022-06 & -0.909 & -1.521 & -1.011 & -1.011 & \textcolor{red}{-0.373} & \underline{-1.778} & -1.301 & \textcolor{red}{\textbf{0.160}} \\
2022-07 & 2.116 & 2.034 & 1.993 & 1.993 & \underline{1.249} & 1.557 & \textcolor{red}{\textbf{2.131}} & 1.797 \\
2022-08 & \underline{-1.290} & \textcolor{red}{-0.745} & \textcolor{red}{-0.833} & \textcolor{red}{-0.833} & \textcolor{red}{\textbf{-0.324}} & \textcolor{red}{-0.797} & \textcolor{red}{-1.091} & \textcolor{red}{-1.184} \\
2022-09 & -1.657 & \textcolor{red}{\textbf{-0.846}} & \textcolor{red}{\textbf{-0.846}} & \textcolor{red}{\textbf{-0.846}} & -1.730 & \underline{-2.628} & -1.847 & \textcolor{red}{-1.269} \\
2022-10 & 0.593 & \textcolor{red}{\textbf{0.761}} & \underline{-0.032} & \underline{-0.032} & 0.429 & \underline{-0.032} & 0.320 & 0.074 \\
2022-11 & \textbf{0.831} & 0.574 & -0.350 & -0.148 & 0.755 & 0.635 & 0.280 & \underline{-0.363} \\
2022-12 & -1.236 & \textcolor{red}{\textbf{-0.613}} & \textcolor{red}{-0.975} & \textcolor{red}{-0.842} & \textcolor{red}{-1.190} & \textcolor{red}{-1.069} & \underline{-1.239} & \textcolor{red}{-0.969} \\
2023-01 & \textbf{1.613} & 1.424 & 1.312 & 1.360 & 0.880 & \underline{0.275} & 1.601 & 1.436 \\
2023-02 & -1.164 & \textcolor{red}{\textbf{-0.058}} & \textcolor{red}{-0.415} & \textcolor{red}{-0.415} & \textcolor{red}{-0.216} & \textcolor{red}{-0.398} & \underline{-1.165} & \textcolor{red}{-0.453} \\
2023-03 & 1.539 & 1.020 & \underline{0.912} & \underline{0.912} & \textcolor{red}{1.552} & 1.507 & \textcolor{red}{\textbf{1.584}} & 1.373 \\
2023-04 & \textbf{0.564} & 0.512 & \underline{0.155} & 0.271 & 0.491 & 0.402 & \textbf{0.564} & 0.549 \\
2023-05 & -0.556 & \underline{-1.281} & -0.906 & -0.711 & \textcolor{red}{\textbf{0.110}} & nan & -0.556 & \textcolor{red}{-0.553} \\
2023-06 & \textbf{0.802} & 0.477 & \underline{-0.243} & 0.324 & 0.416 & nan & \textbf{0.802} & 0.568 \\
2023-07 & \textbf{0.320} & \underline{-0.867} & -0.463 & -0.663 & -0.066 & -0.053 & \textbf{0.320} & -0.154 \\
2023-08 & -0.621 & \textcolor{red}{-0.012} & -1.218 & -1.218 & \underline{-1.970} & -1.173 & -0.621 & \textcolor{red}{\textbf{0.398}} \\
2023-09 & -2.400 & \textcolor{red}{-2.214} & \textcolor{red}{-1.637} & \textcolor{red}{-1.637} & \underline{-2.521} & -2.430 & -2.400 & \textcolor{red}{\textbf{-1.589}} \\
2023-10 & -0.935 & -1.773 & \textcolor{red}{\textbf{-0.376}} & \textcolor{red}{\textbf{-0.376}} & \underline{-2.064} & -1.920 & \textcolor{red}{-0.895} & -1.223 \\
2023-11 & \textbf{2.187} & \underline{1.478} & \underline{1.478} & \underline{1.478} & 1.757 & 1.520 & 1.832 & 1.499 \\
2023-12 & \textbf{1.890} & 1.781 & 1.767 & 1.767 & \textbf{1.890} & \textbf{1.890} & \textbf{1.890} & \underline{0.886} \\
\end{tabular}
\label{tab:strategies_comparison_with_mode-mode_in_sharp-ratio}
\end{table*}

\subsection{Results of Experiment 2: Comparison of Investment Strategies Across Different Time Periods}

Evaluating which strategy performed best on a month-by-month basis, as described in the last subsection, does not fully capture when the LLM-based portfolio management strategies are most effective. To address this, we conducted additional analysis, hypothesizing that the LLM-based strategies perform well during specific periods but not others.

Specifically, we used the US CPI Total indicator to divide the 26-month evaluation period into two types of periods, investigating the best-performing strategies in each. The periods were divided by calculating the 6-month moving average of the year-over-year change in the US CPI Total and determining whether it increased or decreased with respect to the previous month. As a result, the 26 months were categorized as follows: November 2021 to August 2022 and December 2023 were classified as upward trends (High), while September 2022 to November 2023 were classified as downward trends (Low).

We use the following two methods to compare the strategies:
\begin{enumerate} 
\setlength{\parskip}{0cm} 
\setlength{\itemsep}{0cm} 
\item \textbf{Best-mean}: The average value of the target performance metric was calculated during the High or Low period, and the strategy with the best average was selected. 
\item \textbf{Win-ratio-buy-and-hold}: For each month in the High or Low period, the number of months in which a strategy outperformed the buy-and-hold strategy was counted, and the strategy with the highest win ratio was selected. The buy-and-hold strategy itself was excluded from this comparison. 
\end{enumerate}
\noindent Of the four evaluation metrics used, higher values are preferable for return and Sharpe ratio, while lower values are desirable for volatility and maximum drawdown.

Table \ref{tab:strategies_comparison_with_mode-mode_with_CPI} presents the results. From the perspective of the Sharpe ratio, during High periods, the LLM-based strategy pattern 1 is the best strategy according to both the best-mean and win-ratio-buy-and-hold methods. Conversely, in Low periods, pattern 1 is the best in terms of win-ratio-buy-and-hold, but buy-and-hold performs best in terms of best-mean. This suggests that the buy-and-hold strategy may be more suitable during such periods. These results indicate that, in terms of the Sharpe ratio, LLM-based strategies can outperform basic strategies during certain periods, particularly when the CPI trend is upward (i.e., when the 6-month moving average of the year-over-year change in the US CPI Total is rising compared to the previous month).

When considering other evaluation metrics, the results are mixed. For example, in terms of return, the CM(D2) strategy performs best during High periods. For volatility, the RG(S10) strategy often performs best. Regarding maximum drawdown, during High periods, the RG(S5) strategy is the best according to both the best-mean and win-ratio-buy-and-hold methods.

\begin{table}[htbp]
    \centering
    \caption{Best strategies in terms of four metrics (return, volatility, max drawdown and Sharpe ratio) with different CPI trends. }
    \begin{tabular}{cc||c|c}
        Metric & CPI Trend & Best-mean & Win-ratio-buy-and-hold\\
        \hline
        Return & High & Buy-and-hold & CM(D2)\\
         & Low & Buy-and-hold & Pattern 1\\
        \hline
        Volatility & High & Buy-and-hold & RG(S10)\\
         & Low & RG(S10) & RG(S10)\\
        \hline
        Max drawdown & High & RG(S5) & RG(S5)\\
         & Low & Buy-and-hold & Pattern 1\\
        \hline
        Sharpe ratio & High & Pattern 1 & Pattern 1\\
         & Low & Buy-and-hold & Pattern 1\\
        \hline

    \end{tabular}
    \label{tab:strategies_comparison_with_mode-mode_with_CPI}
\end{table}

\begin{figure}[htbp]
    \centering
    \begin{tabular}{c}

        \begin{minipage}{\columnwidth}
            \includegraphics[width=0.95\columnwidth, trim=0 10 0 10, clip]{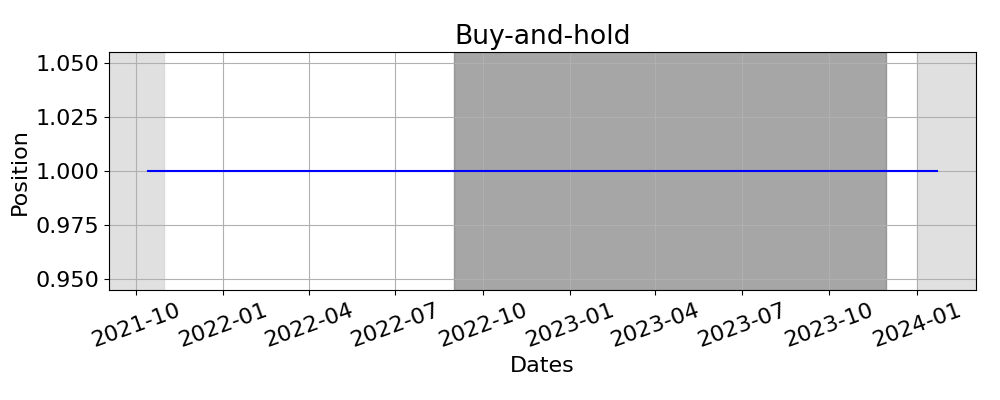}
        \end{minipage}
        \\
        
        \begin{minipage}{\columnwidth}
            \includegraphics[width=0.95\columnwidth, trim=0 10 0 10, clip]{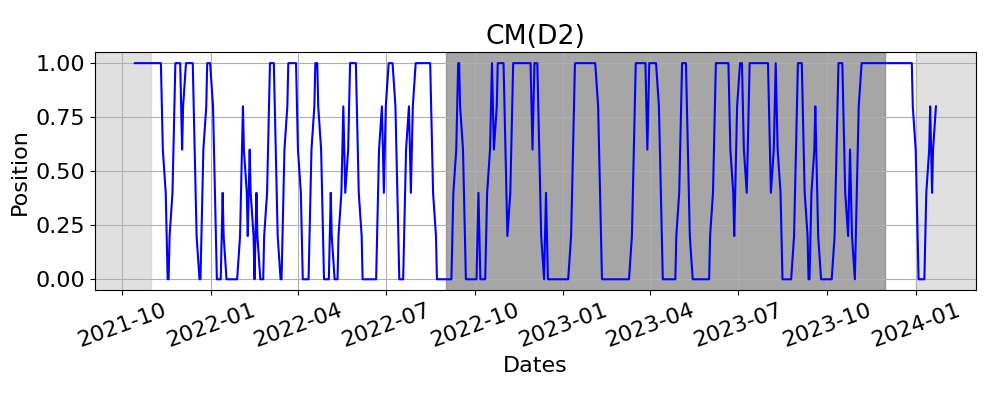}
        \end{minipage}
        \\
        
        \begin{minipage}{\columnwidth}
            \includegraphics[width=0.95\columnwidth, trim=0 10 0 10, clip]{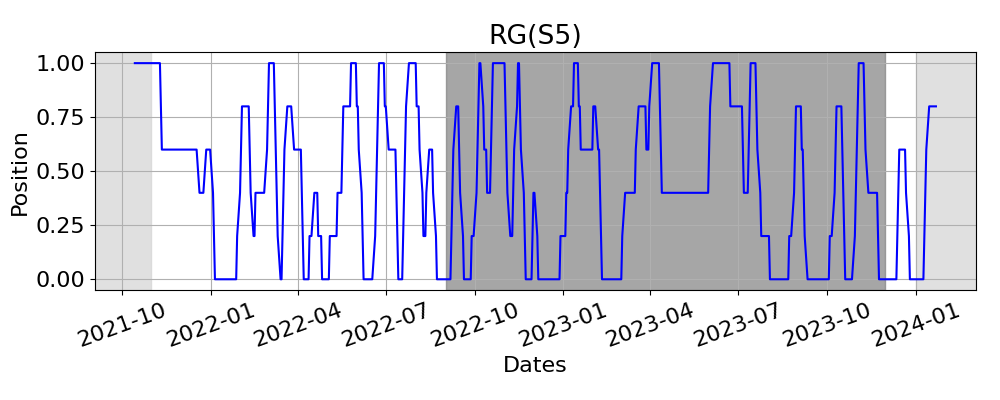}
        \end{minipage}
        \\
        
        \begin{minipage}{\columnwidth}
            \includegraphics[width=0.95\columnwidth, trim=0 10 0 10, clip]{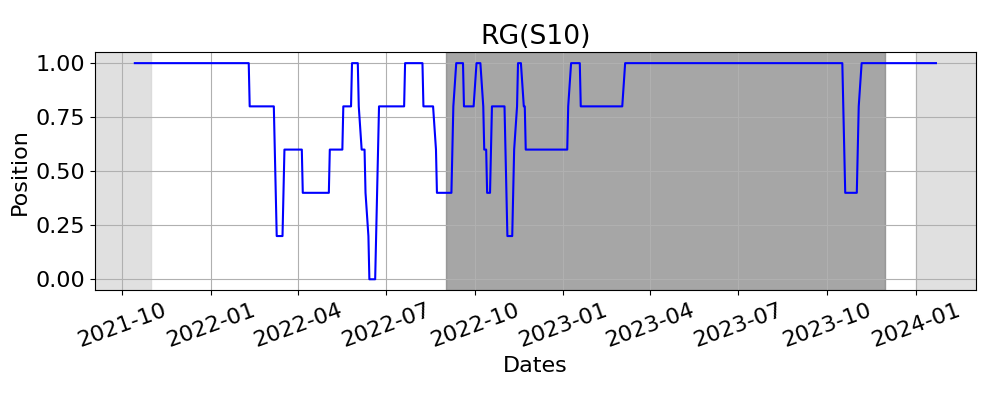}
        \end{minipage}
        \\
        
        \begin{minipage}{\columnwidth}
            \includegraphics[width=0.95\columnwidth, trim=0 10 0 10, clip]{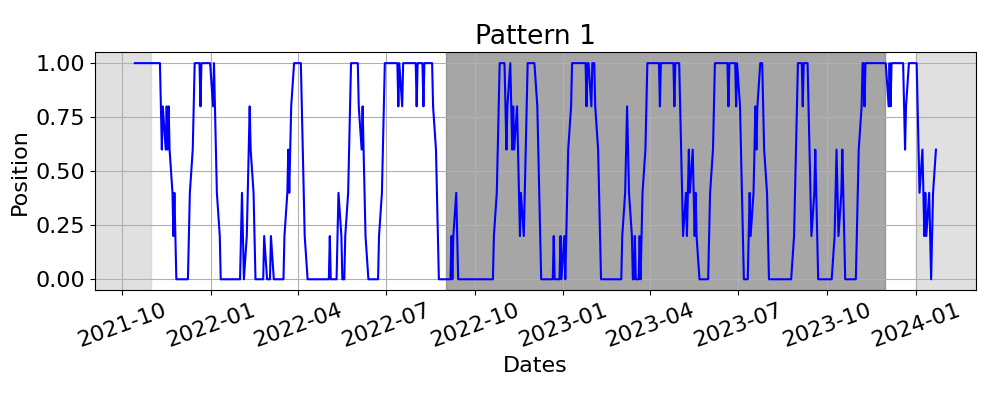}
        \end{minipage}
        \\
        
        \begin{minipage}{\columnwidth}
            \includegraphics[width=0.95\columnwidth, trim=0 10 0 10, clip]{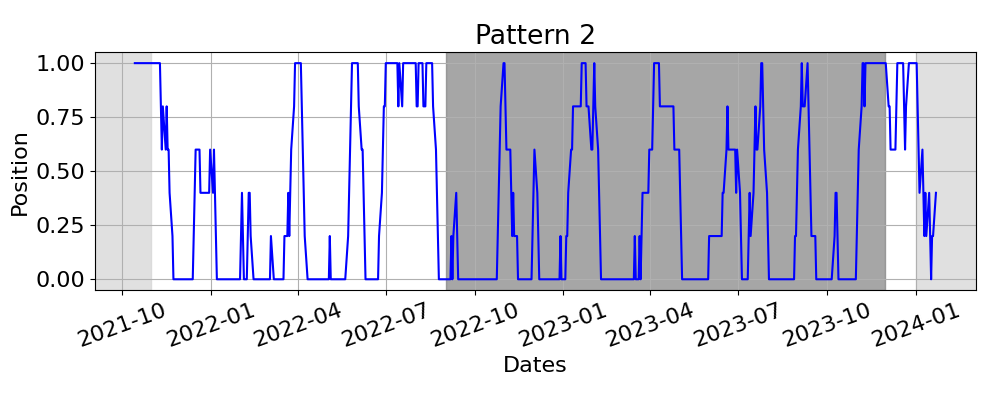}
        \end{minipage}
        \\
        
        \begin{minipage}{\columnwidth}
            \includegraphics[width=0.95\columnwidth, trim=0 10 0 10, clip]{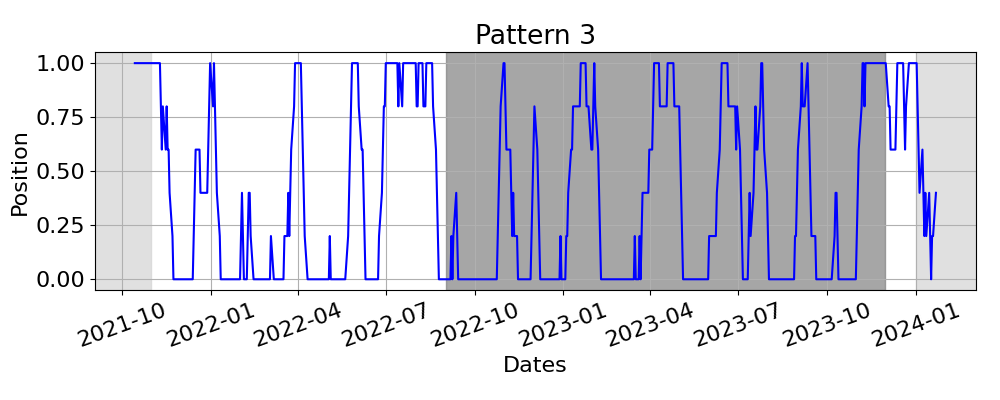}
        \end{minipage}
        \\

    \end{tabular}
    \caption{Changes in the positions of the buy-and-hold, CM(D2), RG(S5), RG(S10), Pattern 1, Pattern 2, and Pattern 3 strategies. The background colors indicate the CPI trend: white for High, dark gray for Low, and light gray for periods outside the evaluation.}
    \label{fig:exp2_plot_position}
\end{figure}

\begin{figure}[htbp]
    \centering
    \begin{tabular}{c}
        
        \begin{minipage}{\columnwidth}
            \includegraphics[width=0.95\columnwidth, trim=0 10 0 10, clip]{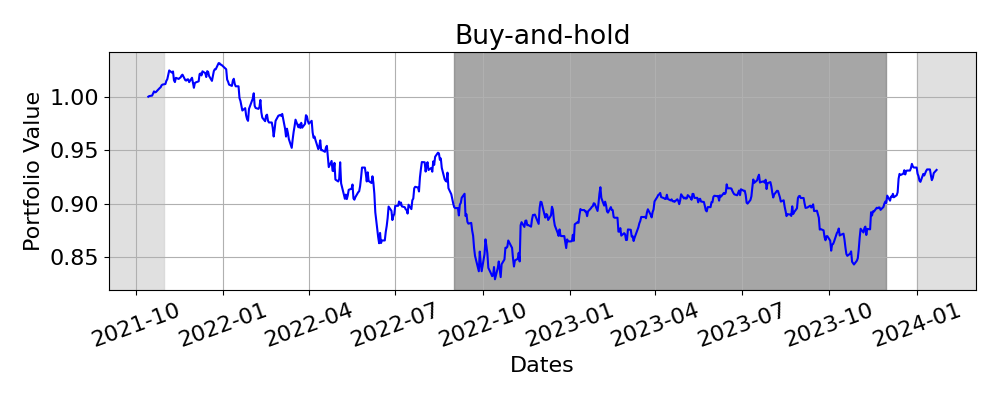}
        \end{minipage}
        \\
        
        \begin{minipage}{\columnwidth}
            \includegraphics[width=0.95\columnwidth, trim=0 10 0 10, clip]{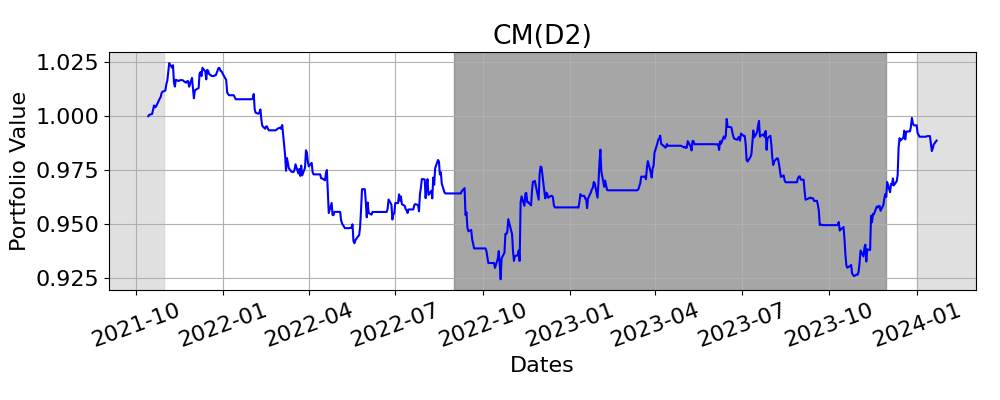}
        \end{minipage}
        \\
        
        \begin{minipage}{\columnwidth}
            \includegraphics[width=0.95\columnwidth, trim=0 10 0 10, clip]{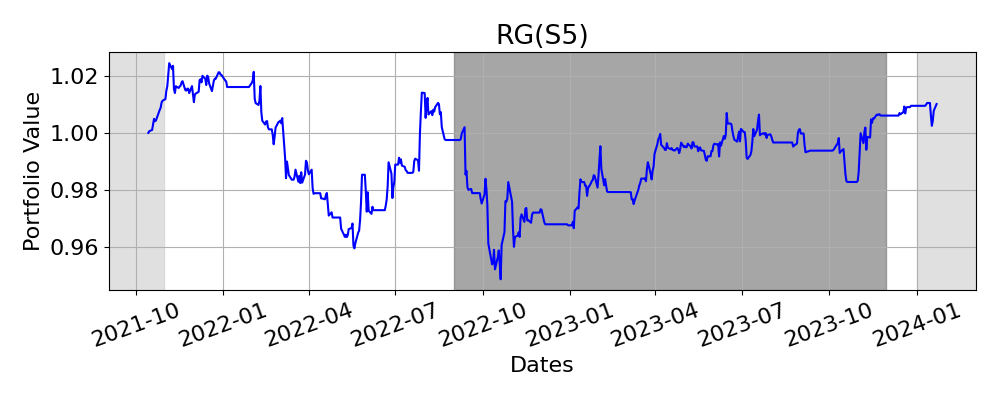}
        \end{minipage}
        \\
        
        \begin{minipage}{\columnwidth}
            \includegraphics[width=0.95\columnwidth, trim=0 10 0 10, clip]{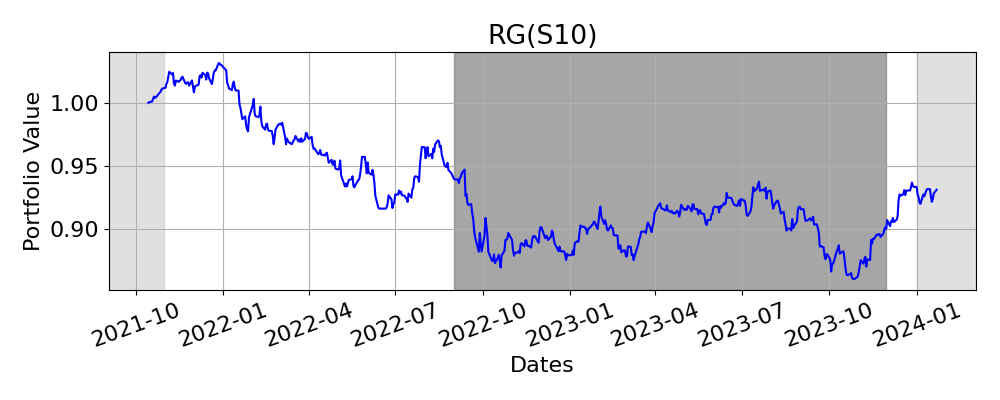}
        \end{minipage}
        \\
        
        \begin{minipage}{\columnwidth}
            \includegraphics[width=0.95\columnwidth, trim=0 10 0 10, clip]{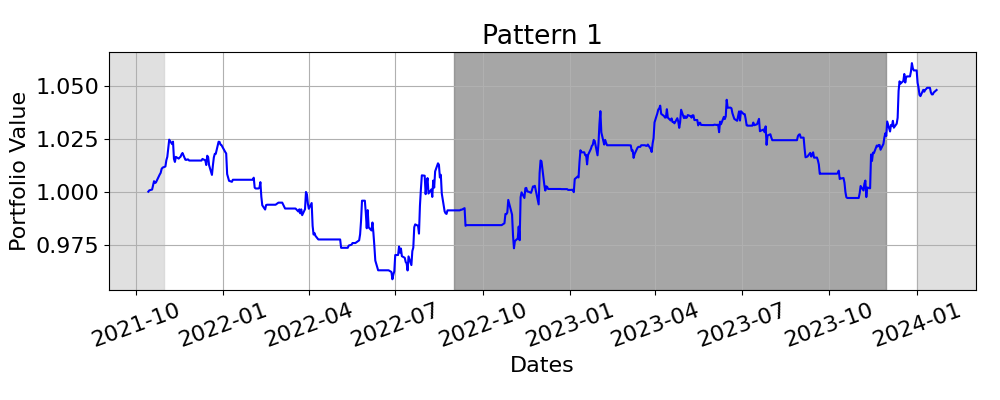}
        \end{minipage}
        \\
        
        \begin{minipage}{\columnwidth}
            \includegraphics[width=0.95\columnwidth, trim=0 10 0 10, clip]{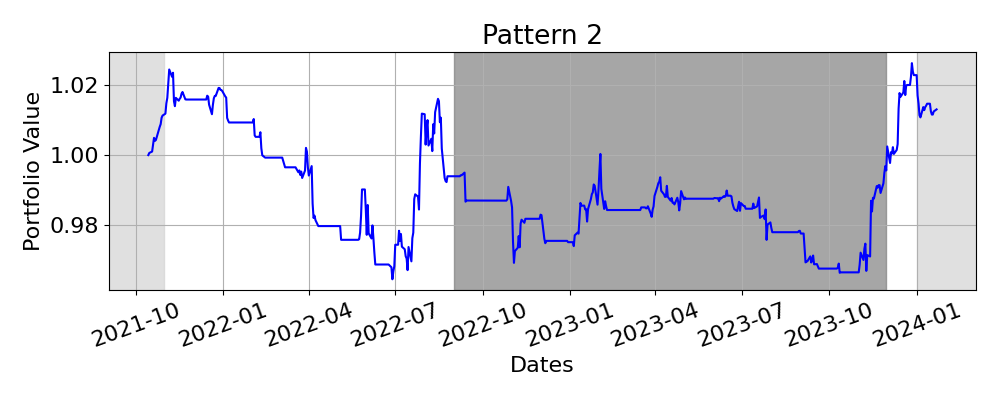}
        \end{minipage}
        \\
        
        \begin{minipage}{\columnwidth}
            \includegraphics[width=0.95\columnwidth, trim=0 10 0 10, clip]{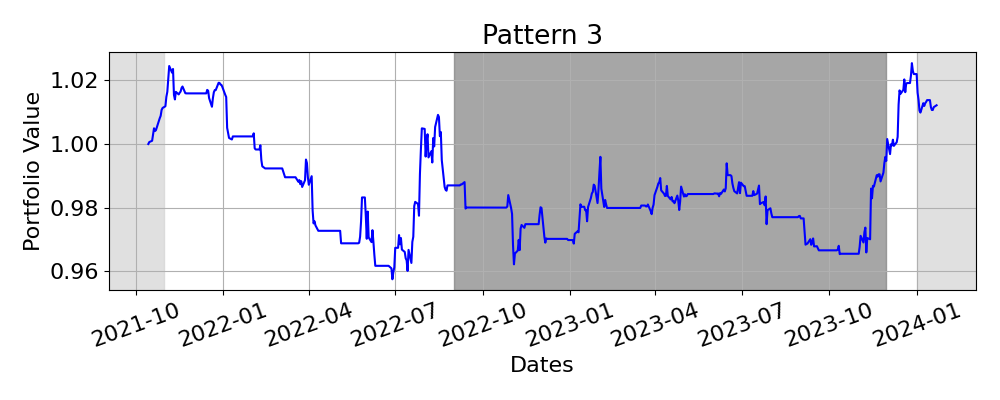}
        \end{minipage}
        \\

    \end{tabular}
    \caption{Changes in the portfolio values of the buy-and-hold, CM(D2), RG(S5), RG(S10), pattern 1, pattern 2, and pattern 3 strategies. The background colors indicate the CPI trend: white for High, dark gray for Low, and light gray for periods outside the evaluation.}
    \label{fig:exp2_plot_portfolio_value}
\end{figure}

To better clarify the differences among strategies, we report the position series for the buy-and-hold, pattern 1, pattern 2, pattern 3, CM(D2), RG(S5), and RG(S10) strategies in Fig. \ref{fig:exp2_plot_position}. Additionally, the portfolio value trends during the investment period (with the initial value set to 1 on the first day) are shown in Fig. \ref{fig:exp2_plot_portfolio_value}. Because LLM predictions often classify movements as class 1 (decline), LLM-based strategies tend to reduce positions. Compared with the pattern 2 and pattern 3 strategies, the pattern 1 strategy increases the position even when class 0 is predicted, allowing the position to return to 1 more easily. This likely facilitates profits during upward trends. The differences between the pattern 2 and pattern 3 strategies were minimal.

Looking at the buy-and-hold strategy in Fig.~\ref{fig:exp2_plot_position}, it appears that there were short-term sharp declines around June 2022, from August to October 2022, around January and March 2023, and again from September to October 2023. During the sharp declines in January and March 2023, not only the LLM-based strategies but also the CM(D2) and RG(S5) strategies reduced their positions to 0, effectively preventing a loss in value. However, during the significant declines from September to October 2022 and September to October 2023, only the LLM-based strategies (patterns 1, 2, and 3) reduced their portfolio positions to 0, thereby avoiding value depreciation. In contrast, during the sharp decline in June 2022, the LLM-based strategies were slow to respond, allowing the value to drop, whereas the CM(D2) and RG(S5) strategies managed to lower their positions to 0, partially mitigating the loss. From these observations, it can be concluded that LLM-based strategies can sometimes detect short-term sharp declines effectively, but there are also instances where basic strategies outperform them in responding to these declines.

Additionally, examining the buy-and-hold strategy from a macroscopic trend perspective, we observe a continuous downward trend during the High period from November 2021 to August 2022. Starting in September 2022, when the CPI Trend shifted to Low, the overall trend remained relatively flat despite some fluctuations (Fig. \ref{fig:exp2_plot_portfolio_value}). Based on these results, we conclude that LLM-based strategies achieve a higher Sharpe ratio during periods of high CPI Trend, particularly when there is a macroscopic downward trend.


\subsection{Qualitative Analysis of the Reasoning of the LLM}

\begin{table}[htbp]
\caption{Differences in reasoning among various personas}
\label{table:reasoning}
\centering
\begin{adjustbox}{max width=0.5\textwidth}
\begin{tabular}{|l|l|l|}
    \hline
    \multicolumn{3}{|c|}{\textbf{Short-term persona}} \\ \hline
    \textbf{Cause} & \textbf{Effect} & \textbf{Share} \\ \hline
    A decline is anticipated & Class 1 is predicted & 4.898 \\ \hline
    Increasing interest rate spread & A decline is anticipated & 2.169 \\ \hline
    The portfolio might face downward pressure & Class 1 is predicted & 1.998 \\ \hline
    Increased market volatility & A decline is anticipated & 1.771 \\ \hline
    Increasing interest rate spread & \makecell[l]{The portfolio might face \\ downward pressure} & 1.300 \\ \hline
    \makecell[l]{Downward trend in the \\ portfolio and stocks} & A decline is anticipated & 0.991 \\ \hline
    Flattening yield curve & A decline is anticipated & 0.958 \\ \hline
    Potential for growth & Class 2 is predicted & 0.853 \\ \hline
    \makecell[l]{General downward trend in the \\ portfolio and futures} & A decline is anticipated & 0.747 \\ \hline
    Increased market volatility & \makecell[l]{The portfolio might face \\ downward pressure} & 0.731 \\ \hline
    Increase in volatility & A decline is anticipated & 0.666 \\ \hline
    Fluctuations in interest rates & A decline is anticipated & 0.471 \\ \hline
    Stronger dollar & \makecell[l]{The portfolio might face \\ downward pressure} & 0.439 \\ \hline
    Strengthening dollar & \makecell[l]{The portfolio might face \\ downward pressure} & 0.431 \\ \hline
    Stronger dollar & A decline is anticipated & 0.431 \\ 
    \hline
    \hline
    \multicolumn{3}{|c|}{\textbf{Medium term persona}} \\ \hline
    \textbf{Cause} & \textbf{Effect} & \textbf{Share} \\ \hline
    A decline is anticipated & Class 1 is predicted & 3.984 \\ \hline
    Increasing interest rate spread & A decline is anticipated & 2.123 \\ \hline
    Increased market volatility & A decline is anticipated & 1.957 \\ \hline
    \makecell[l]{The portfolio might face \\ downward pressure} & Class 1 is predicted & 1.315 \\ \hline
    Flattening yield curve & A decline is anticipated & 0.943 \\ \hline
    Increasing interest rate spread & \makecell[l]{The portfolio might face \\ downward pressure} & 0.919 \\ \hline
    \makecell[l]{Downward trends in the \\ portfolio and stocks} & A decline is anticipated & 0.840 \\ \hline
    potential for growth & 2 & 0.792 \\ \hline
    \makecell[l]{A decline in the portfolio \\ is anticipated} & Class 1 is predicted & 0.784 \\ \hline
    \makecell[l]{General downward trends in the \\ portfolio and futures} & A decline is anticipated & 0.721 \\ \hline
    \makecell[l]{A decline in the portfolio's \\ price is anticipated} & Class 1 is predicted & 0.642 \\ \hline
    Increased market volatility & \makecell[l]{The portfolio might face \\ downward pressure} & 0.578 \\ \hline
    Increasing interest rate spread  & \makecell[l]{A decline in the portfolio's \\ price is anticipated} & 0.483 \\ \hline
    Increased market volatility & \makecell[l]{A decline in the portfolio \\  is anticipated} & 0.452 \\ \hline
    \makecell[l]{General downward trend \\ of the portfolio} & A decline is anticipated & 0.444 \\ \hline
    \hline
    \multicolumn{3}{|c|}{\textbf{Long-term persona}} \\ \hline
    \textbf{Cause} & \textbf{Effect} & \textbf{Share} \\ \hline
    A decline is anticipated & Class 1 is predicted & 4.276 \\ \hline
    Increasing interest rate spread & A decline is anticipated & 2.005 \\ \hline
    Increased market volatility & A decline is anticipated & 1.860 \\ \hline
    Potential for growth & Class 2 predicted & 1.103 \\ \hline
    Flattening yield curve & A decline is anticipated & 1.087 \\ \hline
    \makecell[l]{General downward trends in the \\ portfolio and futures} & A decline is anticipated & 0.894 \\ \hline
    \makecell[l]{The portfolio might face \\ downward pressure} & Class 1 is predicted & 0.886 \\ \hline
    \makecell[l]{Downward trends in the \\ portfolio and stocks} & A decline is anticipated & 0.644 \\ \hline
    Increasing interest rate spread & \makecell[l]{The portfolio might face \\ downward pressure} & 0.612 \\ \hline
    \makecell[l]{Significant price movement \\ is not anticipated} & Class 0 is predicted & 0.596 \\ \hline
    Increase in volatility & A decline is anticipated & 0.499 \\ \hline
    Yield curve steepens & Potential for growth & 0.491 \\ \hline
    A decline in the portfolio is anticipated & Class 1 is predicted & 0.378 \\ \hline
    Potential growth in the portfolio & Class 2 is predicted & 0.370 \\ \hline
    Fluctuations in interest rates & A decline is anticipated & 0.362 \\ \hline
\end{tabular}
\end{adjustbox}
\end{table}

We analyzed the reasoning structure for each persona using the explanations GPT gave for the logic behind its predictions. For instance, when predicting a decline, GPT using a short-term persona stated, ``Rising interest rates and VIX suggest market uncertainty, while a stronger dollar could pressure exports, potentially impacting the portfolio negatively.'' For a rise, GPT using a long-term persona explained, ``Despite market volatility, the overall growth in stock futures and steepening yield curve suggest potential economic growth, which could positively impact the portfolio.'' We extracted the cause-and-effect relationships, ensuring GPT avoided hallucinations using the method in \cite{Matsuoka2024}. Minor phrasing variations were handled via phrase embedding and clustering. The prompt is provided in our repository.

Table \ref{table:reasoning} highlights the top 15 cause-and-effect relationships for each persona. Short-term predictions emphasized declines (class 1: 6.896; class 2: 0.853), driven by factors such as interest rate spreads, market volatility, ``flattening yield curve,'' and the value of the dollar. In the medium term, pessimism persisted (class 1: 6.725; class 2: 0.792), but there was a shift toward stability as the dollar's impact diminished. Long-term predictions reflected growing optimism (class 1: 5.540; class 2: 1.473), with growth factors such as ``potential for growth'' (1.103) and ``yield curve steepening'' causing the ``potential for growth'' (0.491) to gain prominence. This suggests a shift from a decline-focused outlook to a more growth-oriented perspective, with the yield curve evolving from a source of pressure to an indicator of potential growth.

\section{Conclusion}

In this paper, we showed that LLM-based predictions, particularly when combined with the mode ensemble, demonstrated strong performance in detecting market declines. LLM-based investment strategies outperformed the buy-and-hold strategy in terms of Sharpe ratio during periods of upward CPI trends, whereas the buy-and-hold strategy performed better during downward trends. Additionally, other strategies obtained better metrics such as return, volatility, and maximum drawdown, depending on market conditions.

One possible reason for the strong performance of our LLM-based strategies during the CPI uptrend is that baseline strategies, which consider only the past 10 days, rely solely on this limited data to adjust position sizes. This makes them highly sensitive to fluctuations within such a narrow window. In contrast, while our LLM strategy also examines only the last 10 days, it draws on its underlying knowledge to recognize that the recent price movements are part of a larger downward trend. This likely enabled the LLM strategy to make more informed adjustments, scaling down portfolio positions accordingly.

To illustrate, starting in December 2021, the U.S. began to acknowledge that rising inflation was not just a temporary consequence of COVID-19. This shift in understanding led to a series of interest rate hikes, especially rapid in the first year following December 2021. As a result, bond prices entered a range-bound phase, with CPI trends reflecting both upward and downward shifts. These patterns mirrored broader global trends observed in buy-and-hold strategy values, which tended to either decline or oscillate depending on the period.

\section*{Acknowledgment}
This research was supported by JST SPRING GX project (Grant Number JPMJSP2108), the JST FOREST Program (Grant Number JPMJFR216Q) and The University of Tokyo Data Science School. We thank Kimberly Moravec, PhD, from Edanz for editing a draft of this manuscript.

\end{document}